# The color of the Galaxy

Guillermo Abramson[1]
*Statistical and Interdisciplinary Physics Division*
*Centro Atómico Bariloche, CONICET and Instituto Balseiro*
*8402AGP Bariloche, Argentina*

July 25, 2023

*«What color is the universe?»,* asked my friend, while we were making cardboard spectroscopes and testing them with colorful discharge lamps of elementary gases. She is a visual artist interested in everything dealing with colors, and in particular with the colors of astronomical objects. The first thing that came to my mind was that (most) stars emit most of their radiation in the visible range of the electromagnetic spectrum, between 400 and 700 nm. The average of them all came to my mind as a pale grey. What color is the universe? That day, from my balcony, the landscape was indeed monochromatic (Fig. 1).

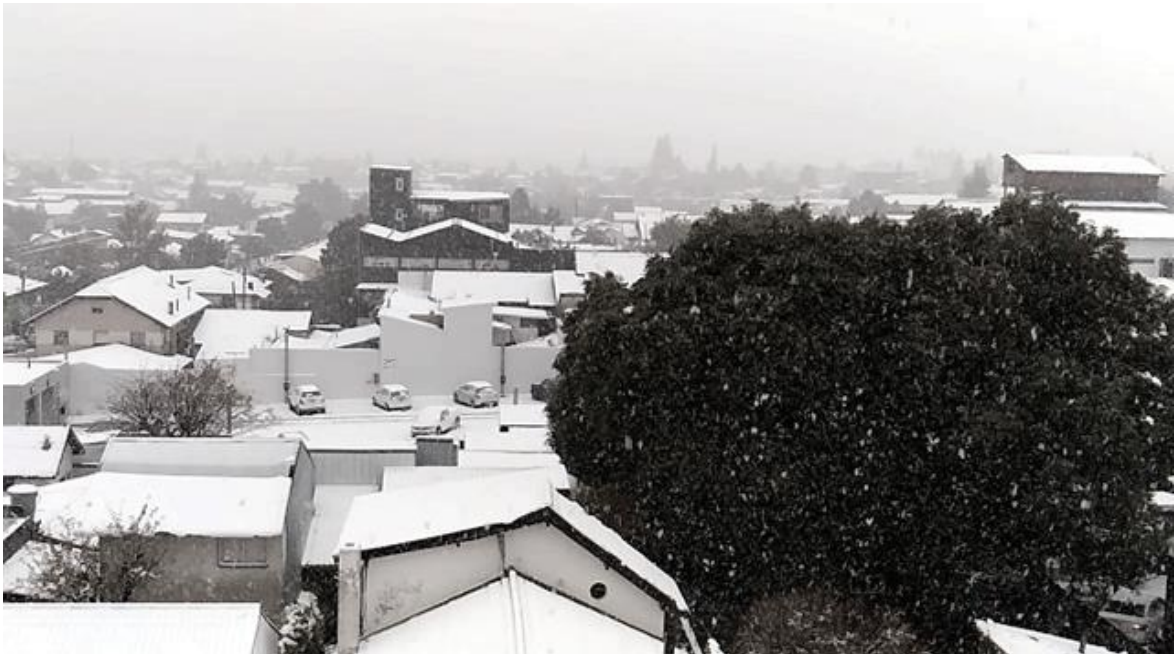

Figure 1. Snow turns all landscapes very monochromatic. (Credit: G Abramson.)

But mine is just one corner of one country on one continent on one planet that's a corner of a galaxy that's a corner of a universe.[2] Astronomical photographs are, instead, so colorful (Fig. 2). Those colors of bright nebulae are indeed saturated, but they are also faint. Astrophotographers tend to emphasize them in the digital processing of the images. But also scientific images of nebulae, while colorful, hide the fact that the so called "bright" nebulae are rather dull. The overwhelming source of light in the universe are stars, stars like the Sun. What color is the Sun? What color are the stars? Could we average them, and find the color of the starry sky?

---

[1] Email: guillermo.abramson@ib.edu.ar.
[2] «But this is one corner of one country on one continent on one planet that's a corner of a galaxy that's a corner of a universe that is forever growing and shrinking and creating and destroying and never remaining the same for a single millisecond, and there is so much, so much, to see.» —The Doctor. Doctor Who, *The Power of Three*, Series 7 Episode 4 (2012).

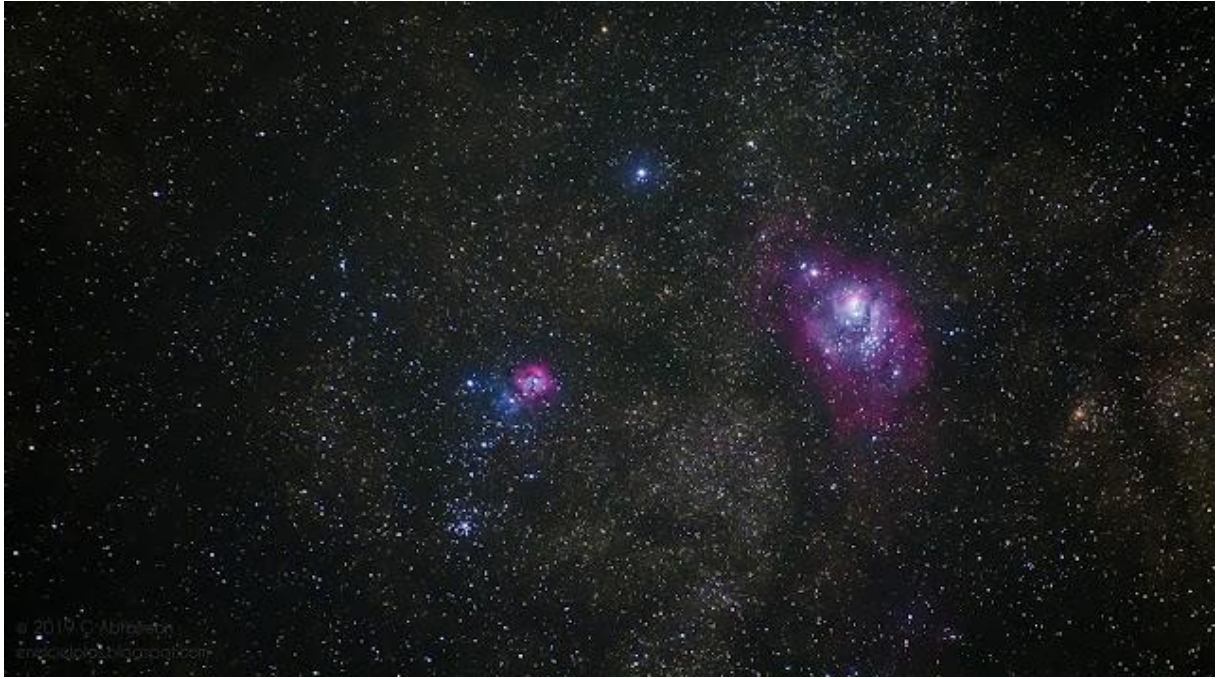

Figure 2. The Lagoon and the Trifid nebulae are large and colorful regions of star formation of the Milky Way galaxy. Their colors are mainly from fluorescence of interstellar gas, irradiated by the young stars that from within them. (Credit: G Abramson.)

To prevent a bias in this endeavor, which could be induced by photographs manipulated with aesthetic criteria, it is best to rely on a good catalog of stellar data. Nowadays, the best such resource is the EDR3 Gaia catalogue [1,2]. Gaia is a space telescope dedicated exclusively to a task tedious like no other: to observe nonstop thousands of millions of stars, registering with unprecedented precision their position, distance, brightness and color. The image shown in Fig. 3 is a composition of the starry sky as seen by Gaia.

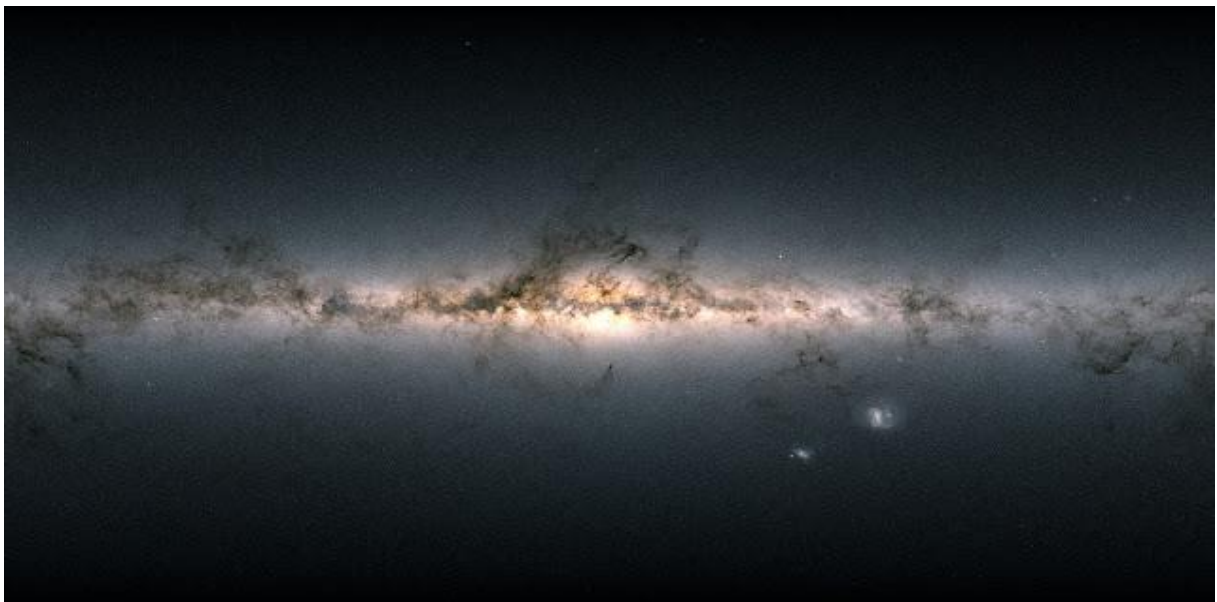

Figure 3. The color of the sky as derived from Gaia EDR3. This is not a photograph, but a graphical representation of the catalogue. (ESA/Gaia/DPAC; CC BY-SA 3.0 IGO. Acknowledgement: A. Moitinho, sci.esa.int/s/WLy3Xp8.)

Figure 3 is not a photograph: it is a graphical representation of the Gaia catalogue, including brightness and color of stars. It is easy to trace the familiar shape of the Milky Way, crisscrossed by dark filaments where cold interstellar dust hides (or reddens) what lies behind. In this image we can see colors: pinkish, orange, whitish, bluish. Can we average them? The most naive thing to try is to average the whole image in an image editing application, such as Photoshop or Gimp. The result is shown in Fig. 4.

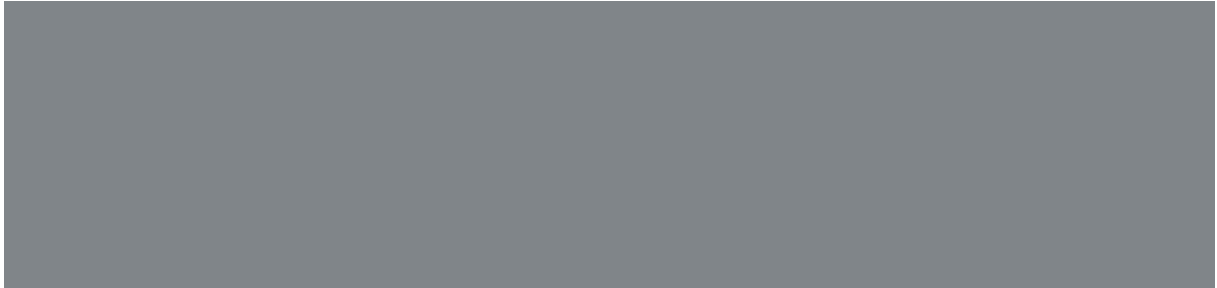
Figure 4. Average of the full resolution (16K) version of the Gaia sky shown in Fig. 3.

This is a barely bluish grey (RGB 128,133,137). It is more or less the color that I imagined when my friend posed her question. But soon I realized that I was averaging the color of the background, which had been arbitrarily set as a dark blue. Was it possible to get rid of it? Again, in the image editing software, I made averages using several brightness masks, to mask-out the dark background. The result appears in Figure 5.

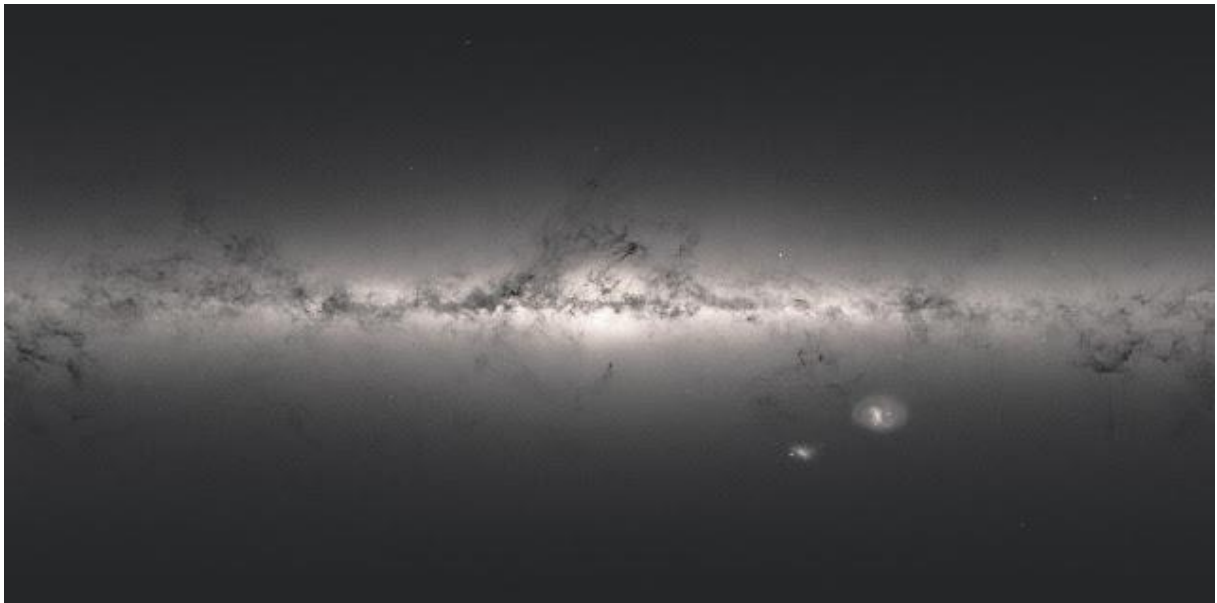
Figure 5. Average of the full resolution (16K) version of the Gaia sky, using a few brightness masks to prevent averaging the dark background.

Now, this was different. We can clearly see the warm tone of the old stars that make up the bulge and the middle stripe of the Milky Way. The overall tone is now beige. The tan galaxy. *The khaki galaxy.*

I was not convinced. Even if I had used the largest version available of this image (16000×8000 pixels), stars are not directly represented, but some kind of average that depends on undisclosed processing done for such a public outreach image. It would be best to rely directly on the actual data. Downloading the more than one billion Gaia stars was out of the question, especially with the bandwidth available in Bariloche. I could use a representative sample, instead. Fortunately, the authors of the catalogue foresaw that such a thing could be useful, and the catalog query allows to download a random sample of the whole.[3] I tried 10000 and the result was nice. I tried a couple of millions and the result was manageable, around 70MB. In the end I downloaded 2.6 million Gaia stars. Of course, I didn't need all the measurements, just brightness and effective temperature. From the temperature one can derive an RGB color using Planck's black body radiation law (a reasonable approximation [3][4]), and then average the three components. I had anticipated a caveat: there are many more red (mainly red dwarves) than blue stars in the Galaxy, but they are much fainter. Could their number outweigh the number of photons towards the red? The right way to proceed was to perform a weighted average of the sources, using brightness as weights. This was done in Mathematica, which has a predefined function that converts temperature into the color of a black body.[5] The result is shown in Figure 6.

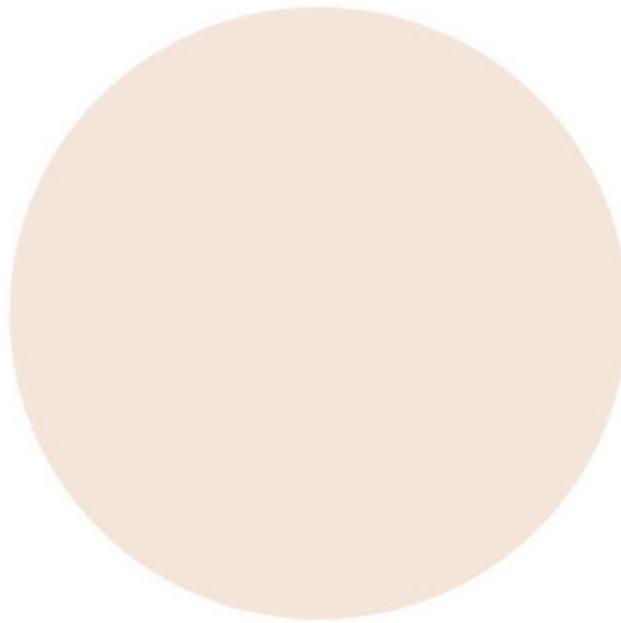

Figure 6. The average color of a random sample of 2.6 million Gaia stars.

---

[3] The following query of the Gaia catalogue produces 2.5E6 sources, some of which have null fluxes, which should be filtered out:
SELECT gaia_source.phot_g_mean_flux,gaia_source.teff_gspphot
FROM gaiadr3.gaia_source
WHERE (gaiadr3.gaia_source.teff_gspphot IS NOT NULL)
AND random_index BETWEEN 0 AND 10000000

[4] A possible alternative to the use of the effective temperature listed in the catalogue (which is not measured, but calculated) would be to use the magnitudes per color band, or the pseudocolor, converting then to RGB. Any of these should require a deeper knowledge of the Gaia documentation, and I do not anticipate a very different result.

[5] If temp is a list of temperatures:
color = ColorData["BlackBodySpectrum", #] & /@ temp;
rgbcolor = Apply[List, #] & /@ color;

This is our result: the color of the Milky Way. The galaxy "tea with milk." In RGB the value is (244,230,219). A gamut without biasing the hue gives a range of chocolate colors shown in Figure 7.

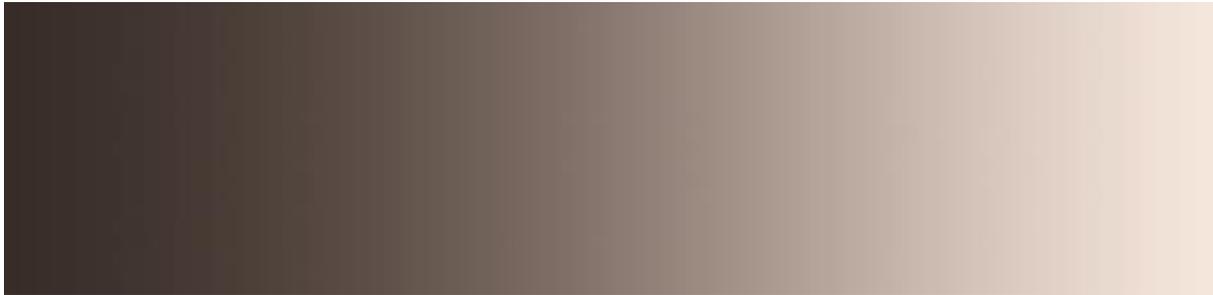

Figure 7. A range of colors, of varying brightness, based on the one shown in Fig. 6,

Of course, I also wondered if this had been done before. I googled "*color of the universe*," and found that a team from Johns Hopkins University had indeed done it 20 years before, but using the colors of 200 thousand galaxies from the Australian 2dF Galaxy Redshift Survey. So, theirs is actually a "color of the universe" instead of a "color of the Galaxy." Their result was also a beige color, very similar to mine, a little less red and more like a champagne. Figure 8 shows both colors, side by side, adjusting brightness for the ease of comparison.

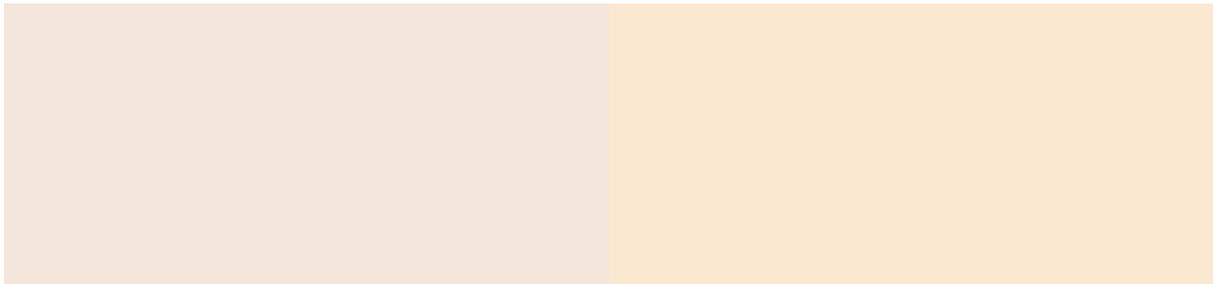

Figure 8. The color of the Milky Way (left, this work) and the color of the universe (right, after [4]).

**Acknowledgements**

The author would like to thank the Isaac Newton Institute for Mathematical Sciences, Cambridge, for support and hospitality during the programme *Mathematics of movement: an interdisciplinary approach to mutual challenges in animal ecology and cell biology,* where work on this paper was undertaken. This work was supported by EPSRC grant no EP/R014604/1. The author also acknowledges the support of ANPCyT through grant PICT 2018-01181, which also helped the visit to Isaac Newton Institute.

This work has made use of data from the European Space Agency (ESA) mission Gaia (www.cosmos.esa.int/gaia), processed by the Gaia Data Processing and Analysis Consortium (DPAC, www.cosmos.esa.int/web/gaia/dpac/consortium). Funding for the DPAC has been provided by national institutions, in particular the institutions participating in the Gaia Multilateral Agreement.